\documentclass[final,5p,times,twocolumn]{elsarticle}

\usepackage{graphicx}

\usepackage{amsmath, amssymb}

\usepackage{color}

\usepackage[colorlinks,bookmarks=false, %
            linkcolor=blue,urlcolor=blue, %
            citecolor=blue]{hyperref}

\biboptions{comma,square,sort&compress}

\makeatletter
\def\ps@pprintTitle{%
 \let\@oddhead\@empty
 \let\@evenhead\@empty
 \def\@oddfoot{}%
 \let\@evenfoot\@oddfoot}
\makeatother

\begin{document}

\begin{frontmatter}



\title{Solving the inverse problem of high numerical aperture focusing using vector Slepian harmonics and vector Slepian multipole fields}


\author{Korn\'el Jahn\corref{cor1}}
\ead{kornel.jahn@gmail.com}
\cortext[cor1]{Corresponding author}
\author{N\'andor Bokor}

\address{Department of Physics, Budapest University of Technology and Economics, 1111 Budapest, Budafoki \'ut 8., Hungary}

\begin{abstract}
A technique using vector Slepian harmonics and vector Slepian multipole fields is presented for a general treatment of the inverse problem of high numerical aperture focusing. A prescribed intensity distribution or electric field distribution in the focal volume is approximated using numerical optimization and the corresponding illuminating field at the entrance pupil is constructed. Three examples from the recent literature are chosen to illustrate the method.
\end{abstract}


\end{frontmatter}


\section{Introduction}

\noindent Engineering the focal spot of a high numerical aperture (NA) lens is a central problem in several optical applications such as laser scanning microscopy, optical trapping and laser micromachining. To obtain the desired spot, one usually has to solve a numerical optimization problem. There exist several examples for the design of annular amplitude or phase masks that tailor a given input beam for a specific application~\cite{zhao2005creation, chen2006three, jabbour2006vector, bokor2007three}. These calculations usually rely on the Debye--Wolf diffraction integral~\cite{wolf1959electromagnetic} to determine the focused field.

Instead of the direct numerical evaluation of the Debye--Wolf integral, some methods use analytical series expansions. These methods include the multipole theory of focusing~\cite{sheppard1997efficient}, the technique of Kant using Gegenbauer polynomials and spherical Bessel functions~\cite{kant1993analytical}, the extended Nijboer--Zernike (ENZ) approach~\cite{braat2003extended} or the scalar eigenfunction expansion of Sherif et al. using Slepian's prolate spheroidal functions~\cite{sherif2008eigenfunction}. Some of these methods were also used for solving the inverse problem of focusing~\cite{kant2000superresolution, braat2005extended, foreman2008inversion}. Recently, a technique using dipole arrays has also been proposed~\cite{wang2010engineering, wang2011three}.

The methods above, however, possess some inherent drawbacks. The scalar Slepian method~\cite{sherif2008eigenfunction} is not suitable for optimization problems with a prescribed intensity pattern~\cite{foreman2008inversion}, the dipole array method is tailored for specific problems~\cite{wang2010engineering, wang2011three} and the multipole theory~\cite{sheppard1997efficient} lacks directionality thus requires additional constraints to ensure directional illumination. Although the ENZ theory~\cite{braat2003extended} is suitable for aberration retrieval where a linearized intensity approximation can be used~\cite{braat2005extended}, it becomes, similarly to the method of Kant~\cite{kant1993analytical}, computationally challenging for arbitrary focal intensity profiles.

Based on the multipole theory of focusing~\cite{sheppard1997efficient}, we have recently proposed an orthonormal basis of vector Slepian harmonics (VSLHs) which are naturally suitable for approximating the illumination in a high NA system~\cite{jahn2012vector}. Its main advantage is that a subset exhibits excellent directionality, i.e. its angular energy distribution is confined to the solid angle of illumination. Each VSLH basis function represents the vectorial plane-wave amplitudes (PWAs) of a focused field described by a corresponding vector Slepian multipole field (VSLMF). The directionality of the VSLHs allows us to approximate common illumination PWAs using a smaller number of coefficients than a representation using the vector spherical harmonics does~\cite{jahn2012vector}.

In this paper, we demonstrate the applicability of these novel vector bases for general inverse problems in high NA focusing. As illustrations, three examples are considered: the optical needle~\cite{wang2010engineering}, tube~\cite{wang2011three} and bubble~\cite{bokor2007three}.

\section{The theory of vector Slepian harmonics and multipole fields}

First we give a brief introduction to the new VSLHs and VSLMFs discussed more comprehensively in Ref.~\cite{jahn2012vector}. One can approximate a divergence-free, monochromatic focused electric field that satisfies the vector Helmholtz equation as
\begin{equation} \label{eq:focused_field_expansion}
	\mathbf{E}(\mathbf{r}) = \sum_{m} \sum_{i=1}^{i_\text{max}} c_{mi} \mathbf{\Phi}_{mi}(\mathbf{r}),
\end{equation}
where the vector functions $\mathbf{\Phi}_{mi}(\mathbf{r})$ are the VSLMFs and $c_{mi}$ denotes complex expansion coefficients (determined in this paper by numerical optimization, as discussed in Section~\ref{sec:inverse}). The VSLMFs themselves are linear combinations of a finite number of vector multipole fields (VMFs) $\mathbf{M}_{lm}(\mathbf{r})$ and $\mathbf{N}_{lm}(\mathbf{r})$~\cite{jahn2012vector} of the same order $m$, i.e.
\begin{equation} \label{eq:vslmf_definition}
	\mathbf{\Phi}_{mi}(\mathbf{r}) := \sum_{l=\ell_m}^L \left[ u_{ilm} \mathbf{M}_{lm}(\mathbf{r}) + v_{ilm} \mathbf{N}_{lm}(\mathbf{r})\right],
\end{equation}
where $L$ denotes the maximal degree of the contributing VMFs and the summation starts at
\begin{equation*}
	\ell_m :=
	\begin{cases}
		1 & \text{if } m = 0 \,,\\
		\lvert m \rvert & \text{otherwise.}
	\end{cases}
\end{equation*}
On the calculation of the coefficients $u_{ilm}$ and $v_{ilm}$ of Eq.~\eqref{eq:vslmf_definition}, see Ref.~\cite{jahn2012vector}.

The azimuthal properties of $\mathbf{\Phi}_{mi}(\mathbf{r})$ are solely determined by the order $m$. In most cases, e.g. for focused fields of linearly or circularly polarized input beams or cylindrical vector beams, only VSLMFs of certain values of $m$ contribute to the focused field. The second integer $i$ indexes VSLMFs of the same order $m$, however, the limit $i_\text{max}$ of Eq.~\eqref{eq:focused_field_expansion} is usually chosen to be much smaller than the number $N_m$ of available VSLMFs for a given order $m$, where
\begin{equation}
	N_m :=
	\begin{cases}
		L & \text{if } m = 0 \,, \\
		L - \lvert m \rvert + 1 & \text{otherwise.}
	\end{cases}
\end{equation}
To clarify the reason for this choice, we must first recall the plane-wave representation of these fields.

The VSLMFs can also be expressed as a superposition of elementary plane-waves, i.e.
\begin{equation} \label{eq:angular_spectrum}
\mathbf{\Phi}_{mi}(\mathbf{r}) = -\frac{\mathrm{i}k}{2\pi} \int\limits_{0}^{2\pi}\int\limits_{0}^{\pi} \mathbf{U}_{mi}(\theta_s, \phi_s) \, \exp(\mathrm{i} k \mathbf{\hat{s}} \cdot \mathbf{r})\,
\sin\theta_s\, \mathrm{d} \theta_s\, \mathrm{d} \phi_s,
\end{equation}
where $k=2\pi/\lambda$ denotes the wavenumber ($\lambda$ is the wavelength), $\theta_s$ and $\phi_s$ are angular spherical coordinates, and $\mathbf{\hat{s}} = \sin \theta_s \cos \phi_s\, \mathbf{\hat{e}}_x + \sin \theta_s \sin \phi_s\, \mathbf{\hat{e}}_y + \cos \theta_s\, \mathbf{\hat{e}}_z$ is the propagation unit vector of a plane-wave component with an amplitude and polarization specified by $\mathbf{U}_{mi}(\theta_s, \phi_s)$. Between $\mathbf{U}_{mi}(\theta_s, \phi_s)$ and $\mathbf{\hat{s}}$ the transversality relation $\mathbf{U}_{mi}(\theta_s, \phi_s) \cdot \mathbf{\hat{s}} = 0$ holds.

Analogously to Eq.~\eqref{eq:vslmf_definition}, the functions $\mathbf{U}_{mi}(\theta, \phi)$ are also defined as a linear combination:
\begin{equation} \label{eq:vslh_definition}
	\mathbf{U}_{mi}(\theta, \phi) := \sum_{l=\ell_m}^L \left[ u_{ilm} \mathbf{Y}_{lm}(\theta, \phi) + v_{ilm} \mathbf{Z}_{lm}(\theta, \phi)\right]\,,
\end{equation}
where $\mathbf{Y}_{lm}(\theta, \phi)$ and $\mathbf{Z}_{lm}(\theta, \phi)$ are vector spherical harmonics (VSHs)~\cite{jahn2012vector}. Since (i) the VSHs represent the PWAs of the VMFs~\cite{devaney1974multipole} similarly to Eq.~\eqref{eq:angular_spectrum} and (ii) VSLMFs are defined as a linear combination of the VMFs according to Eq.~\eqref{eq:vslmf_definition}, the same expansion coefficients $u_{ilm}$ and $v_{ilm}$ appear in Eq.~\eqref{eq:vslh_definition} as in Eq.~\eqref{eq:vslmf_definition}. Furthermore, following from the linearity of our expressions, the combination
\begin{equation}
	\mathbf{A}(\theta_s, \phi_s) = \sum_{m} \sum_{i=1}^{i_\text{max}} c_{mi} \mathbf{U}_{mi}(\theta_s, \phi_s)
\end{equation}
of VSHs with the coefficients $c_{mi}$ of Eq.~\eqref{eq:focused_field_expansion} gives the PWAs $\mathbf{A}(\theta_s, \phi_s)$ of the focused field. The electric field at the entrance pupil of the lens can then be found using the vectorial ray-tracing method of T\"or\"ok et al.~\cite{torok2008high}.

Although Eq.~\eqref{eq:angular_spectrum} resembles the Debye--Wolf integral~\cite{wolf1959electromagnetic} closely, it is important to stress that the integration is performed over the unit sphere of all possible plane-wave directions, not just for the spherical cap
\begin{equation}
	S_\text{C} := \{(\theta_s, \phi_s)\,\vert\, 0 \le \theta_s \le \sin^{-1}(\mathrm{NA}), 0 \le \phi_s < 2\pi\}.
\end{equation}
To be applicable to systems whose NA restricts possible plane-wave directions to $S_\text{C}$, the VSLHs $\mathbf{U}_{mi}(\theta, \phi)$ are constructed to maximize $\eta_{mi}$, the fraction of the energy falling onto $S_\text{C}$~\cite{jahn2012vector}. The VSLHs $\mathbf{U}_{mi}(\theta, \phi)$ for a given $m$ are ordered by decreasing $\eta_{mi}$, thus those with lower $i$ have less ``energy leakage'' outside $S_\text{C}$. In our examples, we restrict ourselves to basis functions with $\eta_{mi} \gtrsim 98\%$. This leads to choosing the maximal index $i_\text{max}$ to be less than $N_m$. We stress the fact that using this method, no extra constraints are necessary to ensure the directionality of the illumination, which would be inevitable when using the multipole theory of focusing~\cite{sheppard1997efficient}.

\section{The inverse problem} \label{sec:inverse}

Next we demonstrate how to use our method of VSLMFs and VSLHs to treat the inverse problem, i.e. to find the illumination for a prescribed focal intensity profile. Here we only consider examples of cylindrical vector beams that can be described as a superposition of radially and azimuthally polarized fields. In the context of VSLHs this means a choice of $m = 0$~\cite{jahn2012vector}. However, all optimization tasks could be performed in a straightforward way for linear, circular or more general polarization states as well, merely with different choices of $m$. Naturally, our method is easily applicable to inverse problems on the electric field distribution, too, provided that a physically feasible focal electric field is prescribed.

In our calculations an aplanatic lens with $\mathrm{NA} = 0.95$ was used. Assuming a cylindrical coordinate system $(\rho,\varphi,z)$ centered at the focus, a rotationally symmetric intensity distribution $I(\rho, z)$ was prescribed in the focal volume.

The focused field can be written as
\begin{equation}
	\mathbf{E}(\mathbf{r}; c_i) = \sum_{i=1}^{i_\text{max}} c_i \mathbf{\Phi}_{0i}(\mathbf{r}),
\end{equation}
where the task is to find the coefficients $c_i$ that minimize the cost function
\begin{equation} \label{eq:cost_function}
	F(c_i) := \iiint_V \left[\mathbf{E}^\ast(\mathbf{r}; c_i) \cdot \mathbf{E}(\mathbf{r}; c_i) - I(\mathbf{r})\right]^2\, \mathrm{d}^3 \mathbf{r},
\end{equation}
the squared error of the intensity integrated over some reasonably chosen volume $V$ around the focus (the asterisk denotes the complex conjugate and we have simplified the notation by omitting the zero $m$ index from $c_{0i}$).

The optimization was performed by sequential least squares programming~\cite{kraft1988software}. Random initial values of $c_i$ were chosen and the process converged to a local minimum of $F(c_i)$. For a given problem, this optimization process was repeated 40 times and the lowest minimum was selected. Although not guaranteed to find a global optimum, the process yielded satisfactory results from a practical point of view. If attaining the global optimum is of crucial importance, computationally intensive global optimizing algorithms can be used.
We note that 40 runs generally yielded only at most 3 different local minima and there was less than 2\% difference in the value of the cost function for the different local minima.

For the examples presented in this paper, we only considered real values of $c_i$ that correspond to a binary phase modulation of the radial and azimuthal components of the input beam. When allowing complex-valued $c_i$ (i.e. a locally elliptically polarized input beam), the size of real optimization parameters doubled and the computation time increased by a factor of $2.4$--$2.7$ on the average. However, we did not obtain any better local optima in terms of cost function value for the examples considered.

The PWAs of $\mathbf{E}(\mathbf{r})$ can be obtained as
\begin{equation}
	\mathbf{A}(\theta_s, \phi_s) = \sum_{i=1}^{i_\text{max}} c_i \mathbf{U}_{0i}(\theta_s, \phi_s).
\end{equation}
Since $\eta_{mi} < 1$~\cite{jahn2012vector}, a small amount of energy leakage may still occur outside $S_\text{C}$. Because of that, the exact focused electric fields were also calculated by Debye--Wolf integration using Chirp Z-Transform~\cite{jahn2010intensity} after enforcing a hard limit at the edge of the entrance pupil.

The prescribed focal intensity functions for the needle, tube and bubble were
\begin{subequations} \label{eq:prescribed_intensities}
\begin{align}
	I(\rho) &\propto \exp\left(-\frac{\rho^2}{a^2}\right), \label{eq:needle_prescribed} \\
	I(\rho) &\propto \frac{\rho^2}{b^2}\, \exp\left(-\frac{\rho^2}{b^2}\right), \label{eq:tube_prescribed} \\
	I(\rho, z) &\propto \frac{\rho^2 + (z/3)^2}{b^2}\, \exp\left[-\frac{\rho^2 + (z/3)^2}{b^2}\right], \label{eq:bubble_prescribed}
\end{align}
\end{subequations}
respectively, where the parameters $a$ and $b$ were chosen such that the full width at half maximum (FWHM) of all functions in the focal plane was $0.4\lambda$. For the optical tube and bubble this value refers to the inner FWHM size of the dark spot. For the optical bubble, a physically feasible aspect ratio $\mathrm{FWHM}_z / \mathrm{FWHM}_\rho = 3$ was chosen, as seen in Eq.~\eqref{eq:bubble_prescribed}. The integration in Eq.~\eqref{eq:cost_function} was performed over a cylinder with a length of $10\lambda$ and a radius of $2\lambda$ for the needle and the tube, and over a prolate ellipsoid of revolution with a length of $12\lambda$ and a radius of $2\lambda$ for the bubble. The shape of the integration volumes was chosen to be similar to each prescribed focal spot; and their lateral size was large enough so that the prescribed intensity at the edges was at least 9 orders of magnitude smaller than the peak value.

In the case of the optical needle and tube, the axial length of the integration volume  approximately matched the maximal length of the focal spot attainable by the VSLMF basis functions involved in the process. This maximal length can be increased by increasing $L$ of Eq.~\eqref{eq:vslmf_definition}~\cite{jahn2012vector}. For all examples presented here, we chose $L=30$.

For both the optical needle and tube, the 11 most concentrated $\mathbf{\Phi}_{0i}(\mathbf{r})$ functions were used. For the needle we included only $\mathbf{\Phi}_{0i}(\mathbf{r})$ corresponding to radially polarized illumination, because the azimuthally polarized components would only increase the lateral FWHM and would not contribute to the on-axis electric field. For analogous reasons, only $\mathbf{\Phi}_{0i}(\mathbf{r})$ corresponding to azimuthally polarized illumination were included for the optical tube. The bubble included both types of polarizations. Moreover, since the ideal optical bubble demands zero electric field in its center, we introduced the extra constraint $\mathbf{E}(\mathbf{r}=\mathbf{0}; c_i) = \mathbf{0}$ in the optimization process.

\section{Results}

\begin{figure}[htb]
	\centering
	\includegraphics{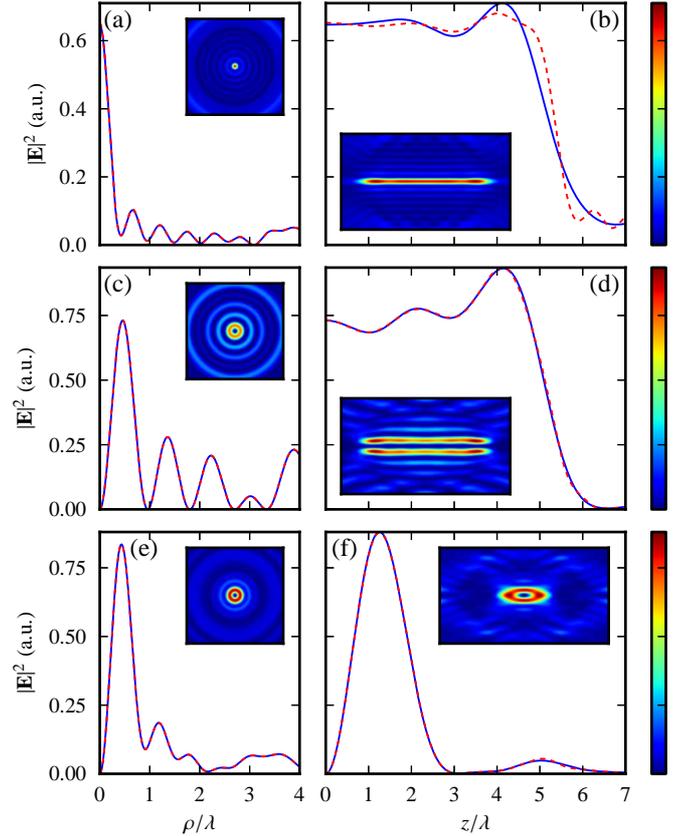}
	\caption{Radial and axial intensity cross-sections of the optical (a)-(b) needle, (c)-(d) tube and (e)-(f) bubble, respectively, with the solid line showing the result of the Debye--Wolf integration and the dashed line the VSLMF approximation (everywhere except in (b) the two curves practically overlap). The insets of show intensity contour maps in (a), (c), (e) the focal plane and (b), (d), (f) the meridional plane, respectively.}
	\label{fig:ff}
\end{figure}

\begin{figure*}[htb]
	\centering
	\hspace*{-0.2in}
	\includegraphics{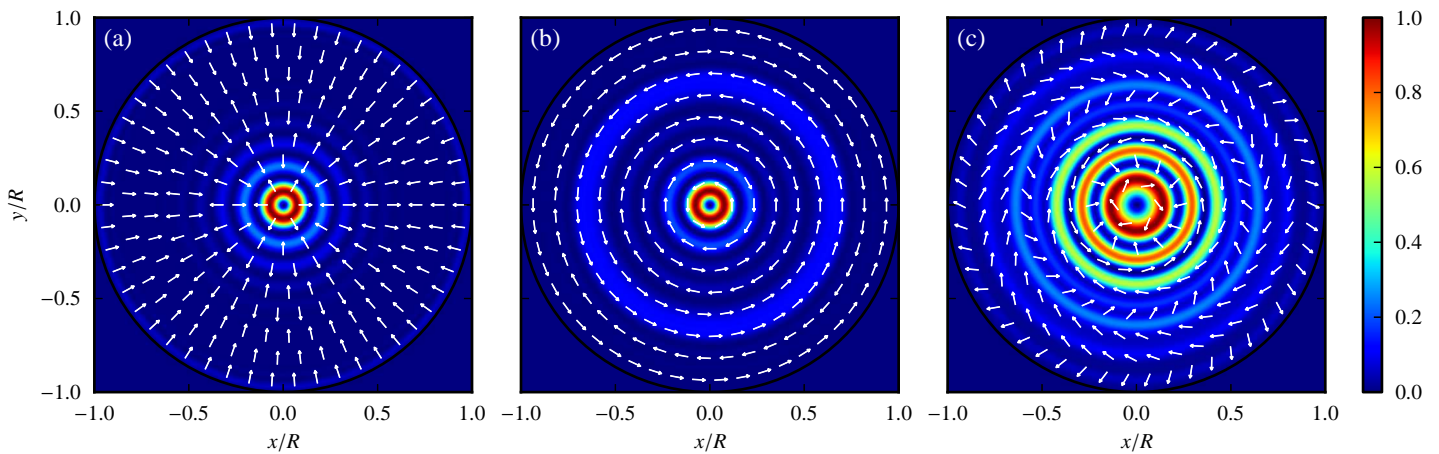}
	\caption{The intensity distribution and polarization of the electric field at the entrance pupil for the optical (a) needle, (b) tube, and (c) bubble ($R$ is the radius of the entrance pupil).}
	\label{fig:as}
\end{figure*}

Fig.~\ref{fig:ff} depicts the intensity distributions obtained from the optimization process for each spot and Fig.~\ref{fig:as} shows the corresponding electric fields at the entrance pupil of the lens. In the case of the needle (Figs.~\ref{fig:ff}(a),(b)), the length of the spot is $9.09\lambda$ (axial full width at 90\% maximum), while the lateral FWHM is $0.41\lambda$ in the focal plane and below $0.44\lambda$ throughout its full length. The beam purity (expressing the fraction of energy of the longitudinally polarized component inside the needle volume, defined in Ref.~\cite{wang2010engineering}) stays above $78\%$ along the needle with a value of $86\%$ in the focal plane. Hence, we can conclude that the quality of the needle resulting from our method is comparable to that of Wang et al.~\cite{wang2010engineering}. Fig.~\ref{fig:ff}(b) shows a slight difference between the Slepian approximation of the focused field using the VSLMFs $\mathbf{\Phi}_{0i}(\mathbf{r})$ and the result of the Debye--Wolf 
integration. This discrepancy can be attributed to the fact that $0.54\%$ of the total energy falls outside the spherical cap of illumination.

The optical tube shown in Figs.~\ref{fig:ff}(c),(d) has a length of $9.09\lambda$, measured along $\rho=\rho_\text{max}$, where $\rho_\text{max}$ is the peak radius in the focal plane. The lateral inner FWHM is $0.46\lambda$ in the focal plane and stays below $0.50\lambda$ throughout the entire length of the tube. Again, our results are comparable with those of Wang et al.~\cite{wang2011three}. Here the energy leakage, when using the Slepian approximation, is only $0.04\%$, since the illumination intensity is low near the edge of the entrance pupil (as seen in Fig.~\ref{fig:as}(b)).

We note that by increasing $L$ the number of total VSLHs increases, including the number of those with $\eta_{mi} \approx 1$. While more optimization parameters are introduced, the maximal attainable length of the tube and needle also increases, as already noted in the previous section. Thus when increased depth of field is of primary importance, a larger value for $L$ can simply be chosen.

Finally, the optical bubble of Figs.~\ref{fig:ff}(e),(f) qualitatively reproduces the result of Bokor and Davidson~\cite{bokor2007three}. In this case, the illumination (Fig.~\ref{fig:as}(c)) is constructed from 22 basis functions. The axial and lateral FWHM of the dark core of the bubble are $1.22\lambda$ and $0.43\lambda$, respectively. As seen in Figs.~\ref{fig:ff}(e),(f), the intensity is zero at the center as prescribed, and the axial and transverse peak intensities are nearly equal in magnitude, which is advantageous e.g. in fluorescence depletion microscopy~\cite{bokor2007three}. Again, the illumination intensity is low near the edge of the entrance pupil, leading to an energy leakage of only $0.03\%$.

\section{Concluding remarks}

We have demonstrated with three well-known examples that VSLHs and VSLMFs are highly suitable for inverse problems of high NA focusing. Knowing the desired 3D intensity distribution in the focal volume, our method can be used effectively to design the electric field distribution at the entrance pupil.

Although the realization of the input field requires careful control of amplitude, phase and polarization, some practical methods using liquid crystal spatial light modulators or form-birefringent spatially variant subwavelength gratings have already been proposed to achieve this task~\cite{wang2007generation, chen2010diffraction, levy2004engineering}.

Finally, it is important to note again that VSLHs and VSLMFs are suitable for treating the inverse problem not only for the focal intensity but for the focal electric field as well. In fact, in that case a least squares approach similar to Eq.~\eqref{eq:cost_function} leads to an easily solvable system of linear equations.

\section*{Acknowledgments}

This work is connected to the scientific program of the ``Development of quality-oriented and harmonized R+D+I strategy and functional model at BME'' project. This project is supported by the New Hungary Development Plan (Project ID: T\'AMOP-4.2.1/B-09/1/KMR-2010-0002).

%








\end{document}